\documentclass[5p]{elsarticle}

\usepackage{graphicx}% Include figure files
\usepackage{dcolumn}% Align table columns on decimal point
\usepackage{bm}% bold math
\usepackage{hyperref}% add hypertext capabilities
\usepackage[T1]{fontenc}
\usepackage[utf8]{inputenc}
\usepackage{pslatex}
\usepackage{textcomp} % for \textdegree
\usepackage{color}
\usepackage{lmodern} % nicer font - The Latin Modern fonts are enhanced versions of the Computer Modern fonts. They have enhanced metrics and glyph coverage.
\usepackage[normalem]{ulem}
\usepackage{amsmath}
\usepackage{amsfonts}
\usepackage{breakurl} %  for breaking urls
\usepackage{array,multirow,graphicx}

\hypersetup{
    breaklinks = true,
    colorlinks = true,
    pdftitle = {}, 
    pdfauthor = {},
    pdfkeywords = {},
    linkcolor = blue,
    citecolor = blue,
    filecolor = black,
    urlcolor = magenta
}

\makeatletter % reset of second corresponding author
\def\@author#1{\g@addto@macro\elsauthors{\normalsize%
    \def\baselinestretch{1}%
    \upshape\authorsep#1\unskip\textsuperscript{%
      \ifx\@fnmark\@empty\else\unskip\sep\@fnmark\let\sep=,\fi
      \ifx\@corref\@empty\else\unskip\sep\@corref\let\sep=,\fi
      }%
    \def\authorsep{\unskip,\space}%
    \global\let\@fnmark\@empty
    \global\let\@corref\@empty  %% Added
    \global\let\sep\@empty}%
    \@eadauthor={#1}
}
\makeatother

\graphicspath{{./figures/}}

\begin{document}

\title{Occupation deficiency in layered structures of UNi$_{x}$Sb$_2$ ($ 0 \leq x \leq 1 $)\\ studied by density functional theory supercell calculations}

\author{M. Werwi\'nski\corref{cor1}}
\cortext[cor1]{Corresponding author} 
\ead{werwinski@ifmpan.poznan.pl}

\author{A. Szajek}

\address{Institute of Molecular Physics, Polish Academy of Sciences,\\ ul. M. Smoluchowskiego 17, 60-179 Pozna\'n, Poland}

\date{\today}

\begin{sloppypar}

\begin{abstract}
The five crystal structures of selected UNi$_{x}$Sb$_2$ compositions are investigated by density functional theory supercell calculations.
The considered phases are USb$_2$, UNi$_{0.33}$Sb$_2$, UNi$_{0.5}$Sb$_2$, UNi$_{0.66}$Sb$_2$, and UNiSb$_2$ ($x$~=~0, 1/3, 1/2, 2/3, 1).
The occupation deficiency of Ni is modeled by removing the Ni layers from constructed supercells followed by relaxation of the structures.
A linear dependence of the lattice parameter $c$ \textit{versus} Ni concentration $x$ is observed, same fulfilling the empirical Vegard's law.
The agreement between results of our calculations with the empirical data from literature confirms the validity of our approach of supercells with empty Ni layers at least in predicting of lattice parameters.
The calculated with orbital polarization magnetic moments on uranium atoms decrease from 1.70~$\mu_{\mathrm{B}}$ to 1.61~$\mu_{\mathrm{B}}$ with increasing Ni concentration $x$.
In comparison to available empirical data of USb$_2$ and UNi$_{0.5}$Sb$_2$, the magnetic moments calculated with orbital polarization are less than 10\% smaller.
\end{abstract}

 \begin{keyword}
Layered structures \sep Uranium compounds \sep \textit{Ab initio} calculations \sep DFT \sep Geometry optimization \sep Supercells
 \end{keyword}

\maketitle

\section{Introduction}
Since the discovery of graphene in 2004 the two-dimensional systems of its analogs have been intensively studied.
Parallel to these efforts also the bulk layered materials have attracted the great attention.
Some of the recent studies focus on the new layered superconductors like BiS$_2$~\cite{mizuguchi_review_2015}, LaXPO (X = Mn, Fe, Ni)~\cite{rahnamaye_aliabad_evaluation_2015}, and (Ca,Pr)FeAs${_2}$~\cite{yakita_new_2014} or the cathode materials for advanced rechargeable Na-ion batteries Na${_x}$MO${_2}$~\cite{su_first-principles_2015, li_electrode_2015}.
The Na${_x}$MO${_2}$ is also an example of a system with occupation deficiency in a layered structure.
Some other samples of layered materials with occupation deficiency are uranium ternaries like UFe$_{0.6}$Sb$_2$~\cite{xie_ferromagnetism_2016} (UFeSb$_2$~\cite{goncalves_crystal_2014}) and UCu$_{0.83}$Sb$_2$~\cite{samsel-czekala_x-ray_2015}.
But it is the UNi$_{0.5}$Sb$_2$ phase which physical properties are studied particularly intensively.~\cite{plackowski_specific_2005, bukowski_electrical_2005, bukowski_single-crystalline_2005, mucha_thermal_2006, davis_hydrostatic_2008,torikachvili_structural_2011, troc_low-temperature_2015}
Some of our previous studies on uranium ternaries addressed similar phases i.e. UCuSb$_2$~\cite{werwinski_first_2014}, UCu$_2$Si$_2$~\cite{morkowski_x-ray_2011}, or URu$_2$Si$_2$~\cite{werwinski_exceptional_2014}.
In this work we investigate the crystal structures of selected UNi$_{x}$Sb$_2$ compositions ($x$~=~0, 1/3, 1/2, 2/3, 1) by density functional theory supercell calculations.
The occupation deficiency of Ni is modeled by a removal of whole Ni layers from constructed supercells which is followed by a relaxation of the structures.
In this paper we will present the calculated crystal structures and compare their parameters to the empirical data from literature.
We will also show the results of the magnetic moments calculations for antiferromagnetic UNi$_{x}$Sb$_2$ structures.
Our method of preparation of the supercells can serve to other experimental and theoretical groups which are dealing with layered materials with occupation deficiency.

\section{\label{sec:details} Details of calculations}
%
%-------------------------------wien2k
%
For \textit{ab initio} calculations we used the full potential linearized augmented plane wave FP-LAPW method as implemented in the WIEN2k code~\cite{peter_blaha_wien2k_2014} %(v 11.1)
with the exchange-correlation potential in the Perdew--Burke--Ernzerhof (PBE) form.~\cite{perdew_generalized_1996}
The spin-orbit coupling does not play an important role in the optimization of the crystal structure, so we included only scalar relativistic effects for the optimization.
In order to get a sufficient accuracy we set the parameters of computations relatively high:
the total energy convergence criterion was $10^{-7}$~Ry,
the plane wave cut-off parameter was \textit{RK$_{max}$}~=~8,
the numbers of \mbox{\textbf{k}-points} in the irreducible wedges of the Brillouin zones were:
256 (16x16x7) for UNiSb$_2$, 110 (20~x~20~x~4) for UNi$_{0.5}$Sb$_2$, 43 (13~x~13~x~2) for UNi$_{0.33}$Sb$_2$ and UNi$_{0.66}$Sb$_2$, and 171 (20~x~20~x~9) for USb$_2$,
the muffin-tin radii were r$_{U}$~=~2.5, r$_{Ni}$~=~2.33, and r$_{Sb}$~=~2.33~a$_0$,
and the force convergence criterion was $10^{-3}$~Ry/a$_0$.
To calculate the magnetic moments we included the spin-orbit coupling term for all orbitals and the orbital polarization corrections~\cite{brooks_calculated_1985, eriksson_orbital_1990} for U~5$f$ and Ni~3$d$ orbitals.
The crystal structures were visualized with VESTA.~\cite{momma_vesta_2008}

%-----------------------------supercell method
%
\begin{figure}
\centering
{\includegraphics[width=0.9\columnwidth]{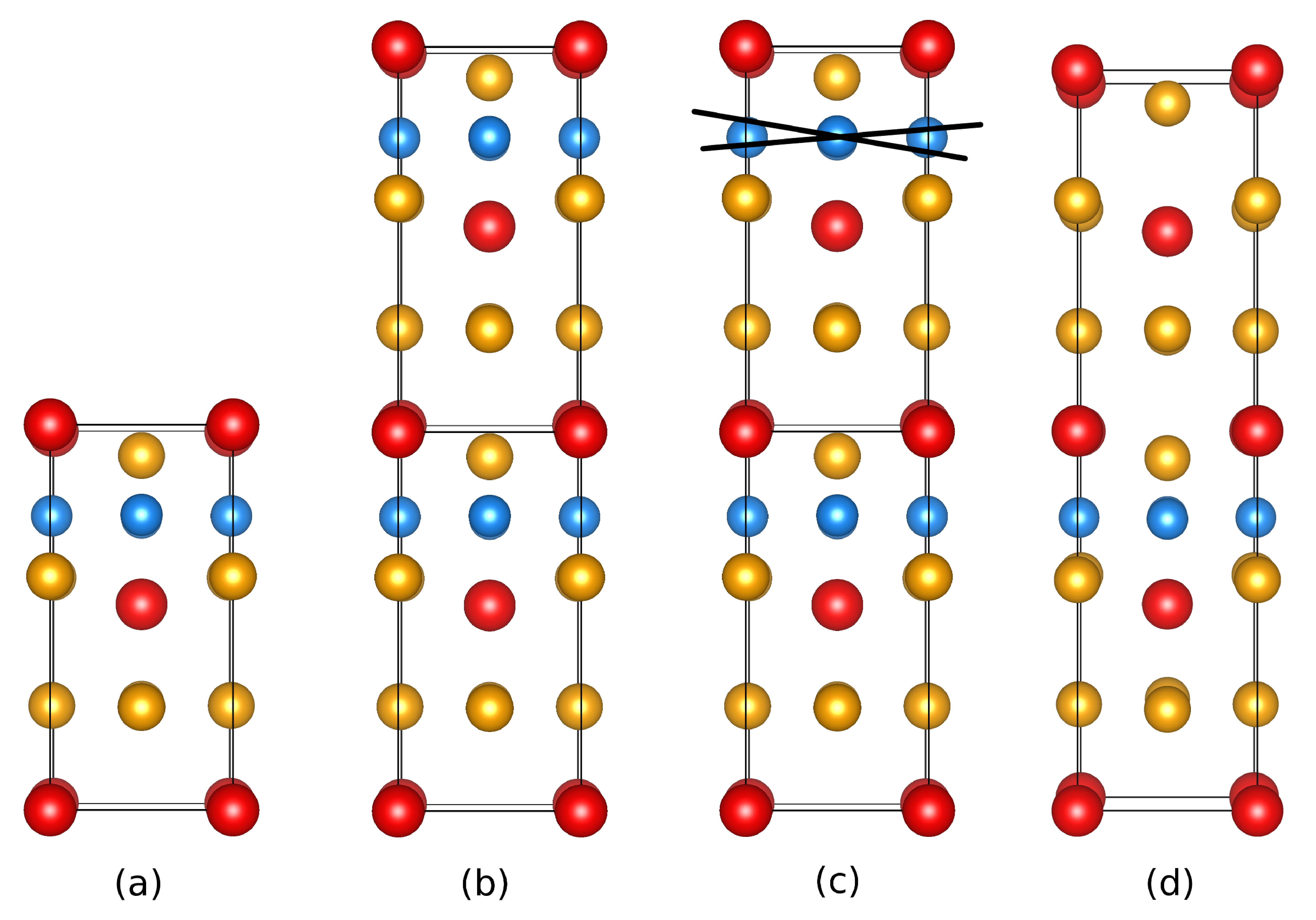}}
\caption{\label{fig:sc_method} 
The four steps of preparation of the supercell model with the occupation deficiency, based on the structure of the UNi$_{0.5}$Sb$_2$ as an example.
a) Starting with a fully occupied single unit cell;
b) multiplication of the fully occupied unit cell along the $z$ direction;
c) removal of the atomic layers in order to reproduce the target concentration;
d) structure optimization by \textit{ab initio} calculations.
The red color is used for U, blue for Ni, and  orange for Sb.
}
\end{figure}

\begin{figure}
\centering
{\includegraphics[width=0.75\columnwidth]{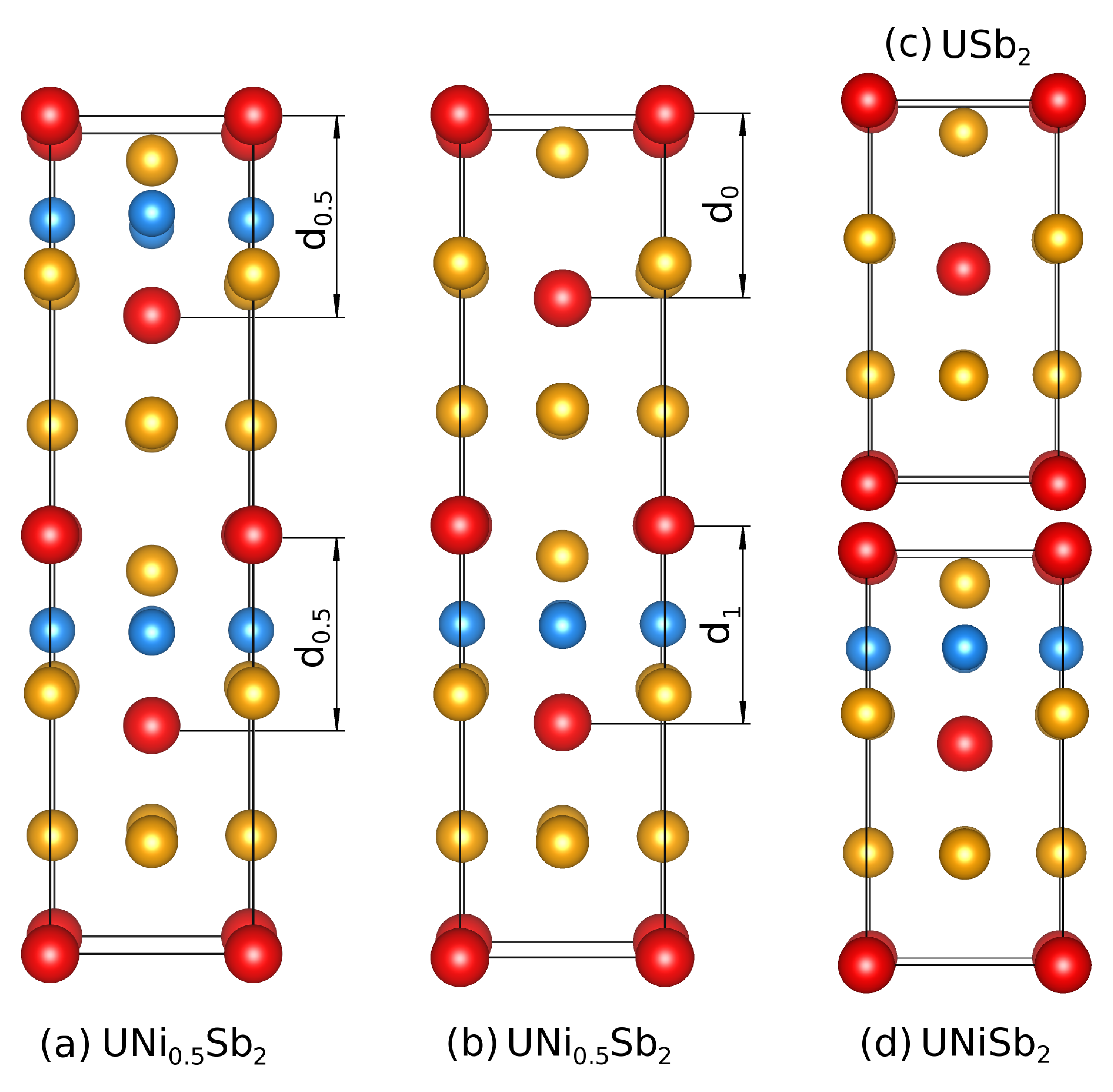}}
\caption{\label{fig:structures} 
The (a) experimentally~\cite{torikachvili_structural_2011} and (b) theoretically predicted crystal structures of UNi$_{0.5}$Sb$_2$. 
(c) USb$_2$ and (d) UNiSb$_2$ crystal structures. 
The red spheres represent U atoms, the blue Ni, and the orange Sb.
In the experimental structure of UNi$_{0.5}$Sb$_2$ (a) every Ni layer is half-occupied. In the proposed model (b) of UNi$_{0.5}$Sb$_2$ one of two Ni layers is removed and the other is fully occupied. The theoretical model (b) of UNi$_{0.5}$Sb$_2$ can be considered as a combination of the USb$_2$ (c) and UNiSb$_2$ (d) crystal structures.}
\end{figure}

%-----------------------------experimental crystallographic data for unisb2 and usb2
%
The preparation of all considered UNi$_{x}$Sb$_2$ crystallographic models starts from the experimental crystallographic data for  a fully occupied unit cell of UNiSb$_2$~\cite{kaczorowski_magnetic_1998,fischer_antiferromagnetism_1989}, see Tab.~\ref{tab:crystal_data_unisb2_usb2} and Fig.~\ref{fig:structures} d).
In Tab.~\ref{tab:crystal_data_unisb2_usb2} we present also the experimental crystal data for USb$_2$~\cite{leciejewicz_neutron-diffraction_1967}.
\begin{table}
\begin{center}
\caption{\label{tab:crystal_data_unisb2_usb2} Experimental crystal data for USb$_2$~\cite{leciejewicz_neutron-diffraction_1967} and UNiSb$_2$.~\cite{kaczorowski_magnetic_1998} Experimental Wyckoff positions z$_{U}$ and z$_{As}$ for UNiAs$_2$ isostructural to  UNiSb$_2$ are z$_{U}$~=~0.2456(4) and z$_{Sb}$~=~0.6531(4).~\cite{fischer_antiferromagnetism_1989}
\vspace{2mm}
}
\begin{tabular}{l|l|l}      % Alignment for each cell: l=left, c=center, r=right
\hline \hline       
 phase    & USb$_2$ 	     	& UNiSb$_2$ 	   \\
\hline 
  sg.     & $P$4/$nmm$ 		&  $P$4/$nmm$          \\
 $a$ (\AA{})  & 4.272 		&  4.322(1)        \\
 $c$ (\AA{})  & 8.741  		&  9.081(1)        \\
 U (2c)   & 1/4 1/4 0.280 	& 1/4 1/4 z$_{U}$  \\
 Sb (2a)  & 1/4 3/4 0  		& 3/4 1/4 0       \\
 Sb (2c)  & 3/4 3/4 0.365  	& 1/4 1/4 z$_{Sb}$ \\
 Ni (2b)  &			& 3/4 1/4 1/2     \\
\hline \hline
\end{tabular} 
\end{center}
\end{table}

The important problem which we had to address is how to model the occupation deficiency in considered UNi$_{x}$Sb$_2$ compositions.
One way would be to use the coherent potential approximation CPA~\cite{edstrom_magnetic_2015} by alloying Ni with empty spheres. 
The other way would be to treat the deficiencies as point defects and prepare the supercells~\cite{reichel_soft_2015} with vacancies, 
as it has been done for an isostructural UCu$_{0.75}$Sb$_{2}$ phase.~\cite{samsel-czekala_x-ray_2015}
However taking into account a linear behavior of $c(x)$ observed for UCu$_{x}$Sb$_2$~\cite{ringe_partial_2008}, we decided to use the supercell approach treating the Ni deficiencies as planar defects. 
Our method can be summarized in four steps: 
a) start with a fully occupied single unit cell (in our case UNiSb$_2$);
b) multiply the fully occupied unit cell along the $z$ direction;
c) remove the atomic layers in order to reproduce the target concentration;
d) use the \textit{ab initio} calculations to optimize the structures.
A sketch of our method is also presented in Fig.~\ref{fig:sc_method}.

\section{Results and discussion} 
\begin{figure}
\centering
{\includegraphics[trim = 110 10 10 40,clip,height=1\columnwidth,angle=270]{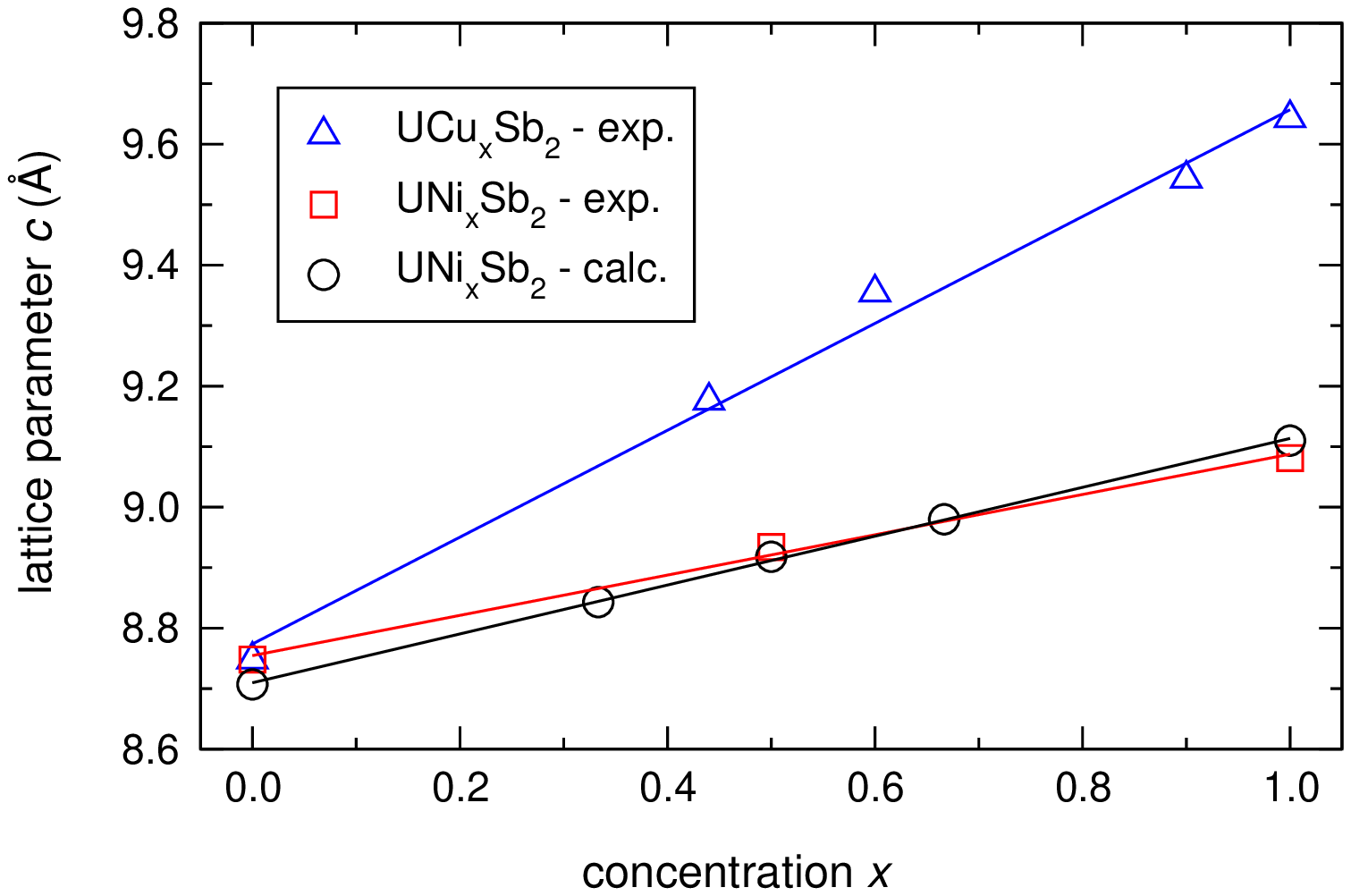}}
\caption{\label{fig:c_vs_x} 
The calculated dependence of lattice parameter $c$ \textit{versus} Ni concentration $x$ for UNi$_{x}$Sb$_2$ system. 
For comparison the corresponding experimental results for UNi$_{x}$Sb$_2$~\cite{leciejewicz_neutron-diffraction_1967, torikachvili_structural_2011, kaczorowski_magnetic_1998} and for  UCu$_{x}$Sb$_2$~\cite{leciejewicz_neutron-diffraction_1967, bobev_uranium_2006, ringe_partial_2008, bukowski_single-crystalline_2005, kaczorowski_magnetic_1998} systems. 
The experimental results for UCu$_{x}$Sb$_2$ have been for the first time collected by Ringe and Ibers.~\cite{ringe_partial_2008}
Symbols are for data points and straight lines for linear regression. 
}
\end{figure}

We noticed that the lattice parameter $c$ \textit{versus} Cu concentration $x$ for the UCu$_{x}$Sb$_2$ system is linear obeying the Vegard's law~\cite{ringe_partial_2008} when the lattice parameter $a$ \textit{versus} $x$ remains more or less the same, see Fig.~\ref{fig:c_vs_x}.
It led us to the assumption that the Ni deficiency in layered UNi$_{x}$Sb$_2$ system takes the form of vacating full Ni layers perpendicular to a tetragonal axis.
We then proposed the simplified model of UNi$_{0.5}$Sb$_2$ which consists of two combined unit cells, one with a full Ni layer and the other with an empty one, 
which looks like a combination of UNiSb$_2$ and USb$_2$ unit cells, see Fig.~\ref{fig:sc_method}~c).
A further support for our reasoning comes from the experimental data, where the sum of USb$_2$ and UNiSb$_2$ lattice parameters $c$ $8.741 + 9.081 = 17.822$~(\AA{})~\cite{leciejewicz_neutron-diffraction_1967, kaczorowski_magnetic_1998}  is very close to the value of $c$~=~17.868~\AA{} for UNi$_{0.5}$Sb$_2$.~\cite{torikachvili_structural_2011}
Based on the above arguments we decided to model the UNi$_{x}$Sb$_2$ layered compounds with occupation deficiency as composed of unit cells with completely full or empty Ni layers in right proportion to Ni concentration.
\begin{table*}
\caption{\label{tab:crystal_data} The crystallographic data of theoretically predicted crystal structures for USb$_2$, UNi$_{0.33}$Sb$_2$, UNi$_{0.5}$Sb$_2$, UNi$_{0.66}$Sb$_2$, and UNiSb$_2$ compounds.}
%\rotatebox{90}{
\begin{tabular}{ll|ll|ll|ll}      % Alignment for each cell: l=left, c=center, r=right
\\
\hline \hline
phase&UNi$_{0.33}$Sb$_2$& phase & UNi$_{0.5}$Sb$_2$ & phase & UNi$_{0.66}$Sb$_2$& phase & UNiSb$_2$ \\ 
\hline
sg.          & P4mm   & sg.         & P4mm  & sg.          & P4mm  &sg.         &  P4/nmm   \\
$a$ (\AA{})  & 4.32   & $a$ (\AA{}) & 4.33 & $a$ (\AA{})  & 4.32  & $a$ (\AA{})&   4.32   \\
$c$ (\AA{})  & 26.60  & $c$ (\AA{}) & 17.84 & $c$ (\AA{})  & 26.95 & $c$ (\AA{})&   9.11   \\ 
U1  & 0.75 0.75 0.235 & U1  & 0.0 0.0 0.000& U1  & 0.75 0.75 0.248& U(2c)  & 0.25 0.25 0.267\\
U2  & 0.75 0.75 0.582 & U2  & 0.5 0.5 0.275& U2  & 0.75 0.75 0.571& Ni(2b) & 0.75 0.25 0.500 \\
U3  & 0.75 0.75 0.908 & U3  & 0.0 0.0 0.513& U3  & 0.75 0.75 0.909& Sb1(2a)& 0.75 0.25 0.000 \\
U4  & 0.25 0.25 0.092 & U4  & 0.5 0.5 0.788& U4  & 0.25 0.25 0.091& Sb2(2c)& 0.25 0.25 0.660 \\ \cline{7-8}
U5  & 0.25 0.25 0.419 & Sb1 & 0.0 0.5 0.137& U5  & 0.25 0.25 0.429& phase & USb$_2$   \\ \cline{7-8}
U6  & 0.25 0.25 0.766 & Sb2 & 0.0 0.0 0.312& U6  & 0.25 0.25 0.752& sg.          & P4/nmm    \\
Sb1 & 0.25 0.25 0.210 & Sb3 & 0.5 0.5 0.476& Sb1 & 0.25 0.25 0.224& $a$ (\AA{})  & 4.326     \\
Sb2 & 0.75 0.75 0.117 & Sb4 & 0.5 0.0 0.651& Sb2 & 0.75 0.75 0.114& $c$ (\AA{})  & 8.707     \\
Sb3 & 0.25 0.25 0.556 & Sb5 & 0.0 0.0 0.824& Sb3 & 0.25 0.25 0.546& U1(2c)  & 0.25 0.25 0.281 \\
Sb4 & 0.75 0.75 0.445 & Sb6 & 0.5 0.5 0.964& Sb4 & 0.75 0.75 0.454& Sb1(2a) & 0.25 0.75 0.000 \\
Sb5 & 0.25 0.25 0.884 & Ni1 & 0.5 0.0 0.394& Sb5 & 0.25 0.25 0.886& Sb2(2c) & 0.75 0.75 0.358 \\
Sb6 & 0.75 0.75 0.791 &     &               & Sb6 & 0.75 0.75 0.776&         &              \\
Sb7 & 0.25 0.75 0.000 &     &               & Sb7 & 0.25 0.75 0.000&         &              \\
Sb8 & 0.25 0.75 0.327 &     &               & Sb8 & 0.25 0.75 0.339&         &              \\
Sb9 & 0.25 0.75 0.674 &     &               & Sb9 & 0.25 0.75 0.661&         &              \\
Ni1 & 0.25 0.75 0.500 &     &               & Ni1 & 0.25 0.75 0.168&         &              \\
    &                 &     &               & Ni2 & 0.25 0.75 0.832&         &              \\
\hline \hline   
 \end{tabular}
% }
\end{table*}
The main result of this work are five crystal structures predicted for USb$_2$, UNi$_{0.33}$Sb$_2$, UNi$_{0.5}$Sb$_2$, UNi$_{0.66}$Sb$_2$, and UNiSb$_2$.
The structures data are collected in Tab.~\ref{tab:crystal_data}.

%----------------------------------\subsection{UNiSb$_2$}
%
We start the calculations from relaxation of the experimental unit cell of UNiSb$_2$~\cite{kaczorowski_magnetic_1998,fischer_antiferromagnetism_1989}, see Tab.~\ref{tab:crystal_data_unisb2_usb2} and Fig.~\ref{fig:structures} d).
The UNiSb$_2$ unit cell contains a single Ni layer perpendicular to the $c$ axis.
The fact that the UNiSb$_2$ is an antiferromagnet (AFM)~\cite{kaczorowski_structural_1992} has to be taken into account in order to get an accurate result of the relaxation.
We do this by lowering the symmetry of the unit cell of UNiSb$_2$ and considering two inequivalent U sites as carrying the opposite magnetic moments.
The lattice parameter $c$~=~9.110~\AA{} calculated from AFM UNiSb$_2$ structure agrees much better with the experimental value $c$~=~9.081~\AA{}~\cite{kaczorowski_magnetic_1998} than the theoretical non-magnetic result $c$~=~8.991~\AA{}.
We conclude that for the UNi$_{x}$Sb$_2$ phases the relaxation should be conducted in AFM spin-polarized mode.

%----------------------------------\subsection{USb$_2$}
%
The first structure calculated by us based on the supercell method is the USb$_2$ structure, see Fig.~\ref{fig:structures}~c) and Tab.~\ref{tab:crystal_data}.
USb$_2$ can be considered as an extreme member of the UNi$_{x}$Sb$_2$ family with the Ni occupation equal to zero.
To model it we have removed the only Ni layer from UNiSb$_2$ single unit cell and optimized the structure.
After the AFM relaxation performed with no symmetry we determined for USb$_2$ the P4/nmm space group, consistent with the experiment.
The fact that the calculated space group, lattice parameters, and Wyckoff positions z$_U$ and z$_{Sb}$ are all in excellent agreement with measured values (see Tables~\ref{tab:crystal_data}~and~\ref{tab:crystal_data_unisb2_usb2}) confirms the validity of the introduced supercell method.

%------------------------------\subsection{UNi$_{0.5}$Sb$_2$}
%
The second layered structure with occupation deficiency calculated with our method is the structure of the UNi$_{0.5}$Sb$_2$, see Tab.~\ref{tab:crystal_data}. 
%
%-----------physical properties
%
The UNi$_{0.5}$Sb$_2$ is an antiferromagnet with the N\'{e}el temperature $T_N$~=~161~K, its magnetic moments are localized on uranium atoms and oriented along the $c$ axis~\cite{bukowski_single_2004} with ordering sequence $+-+-$.~\cite{torikachvili_structural_2011}
The UNi$_{0.5}$Sb$_2$ structure was described in literature twice,
the first approach~\cite{bukowski_single_2004} is based on a single UNiSb$_2$-type unit cell and describes the Ni occupation as close to 0.5.
%
%--------------------torikachvili
%
The second approach by Torikachvili et al.~\cite{torikachvili_structural_2011} is based on a doubled UNiSb$_2$-type unit cell 
containing two half-filled Ni layers, see Fig.~\ref{fig:structures}~a).
In this picture the space group is P4$_2$/nmc and the lattice parameters following $a$~=~4.333(6)~\AA{} and $c$~=~17.868(4)~\AA{}
and the interlayer U-U distance d$_{0.5}$~=~4.15~\AA{}.
The crystal structure of UNi$_{0.5}$Sb$_2$ generated with our supercell method has a space group P4mm and lattice parameters $a$~=~4.336~\AA{} and $c$~=~17.837~\AA{}, see Fig.~\ref{fig:structures}~b). 
Although we observe a very good consistency between lattice parameters from the experimental and theoretical pictures, the corresponding two sets of atomic positions significantly differ. 
In our picture the top Ni layer is emptied and the bottom one is full, see Fig.~\ref{fig:structures}~b).
The calculated interlayer U-U distances d$_{0}$~=~3.79~\AA{} and d$_{1}$~=~4.25~\AA{}, for empty and full Ni layers, correspond well to the U-U distances for USb$_2$ and UNiSb$_2$ (d$_{USb_2}$~=~3.82~\AA{} and d$_{UNiSb_2}$~=~4.24~\AA{}). 
It shows how close our picture of UNi$_{0.5}$Sb$_2$ is to the one combined of USb$_2$ and UNiSb$_2$.

%----------------------------------\vegard
%
As we can see in Fig.~\ref{fig:c_vs_x}, the calculated linear behavior of $c(x)$ for UNi$_{x}$Sb$_2$ phase diagram is in excellent agreement with experimental data from literature.
Linear regression analysis leads to the formula $c = 8.71 + 0.40 x$ (\AA{}) for $c(x)$  calculations for UNi$_{x}$Sb$_2$ and $c = 8.75 + 0.33 x$ (\AA{}) for corresponding experimental data.
For isostructural UCu$_{x}$Sb$_2$ also the linear behavior is observed following a formula $c = 8.77 + 0.88 x$  (\AA{}).
The above result can be also compared with a relation found experimentally for a martensite with a tetragonal distortion $c_{FeC} = 2.87 + 0.116 x_C$ (\AA{}), where the lattice parameter $c$ is proportional to the amount of carbon $x_C$ in the alloy (for small C concentrations).~\cite{kurdjumov_nature_1975, delczeg-czirjak_stabilization_2014}
The calculated linear behavior of $c(x)$ for UNi$_{x}$Sb$_2$ is a strong evidence that our approach is a suitable method of modeling the layered structures with occupation deficiency.
It can be used both to improve structure determination with XRD and as an independent theoretical tool.

\begin{figure}
\centering
{\includegraphics[width=0.99\columnwidth]{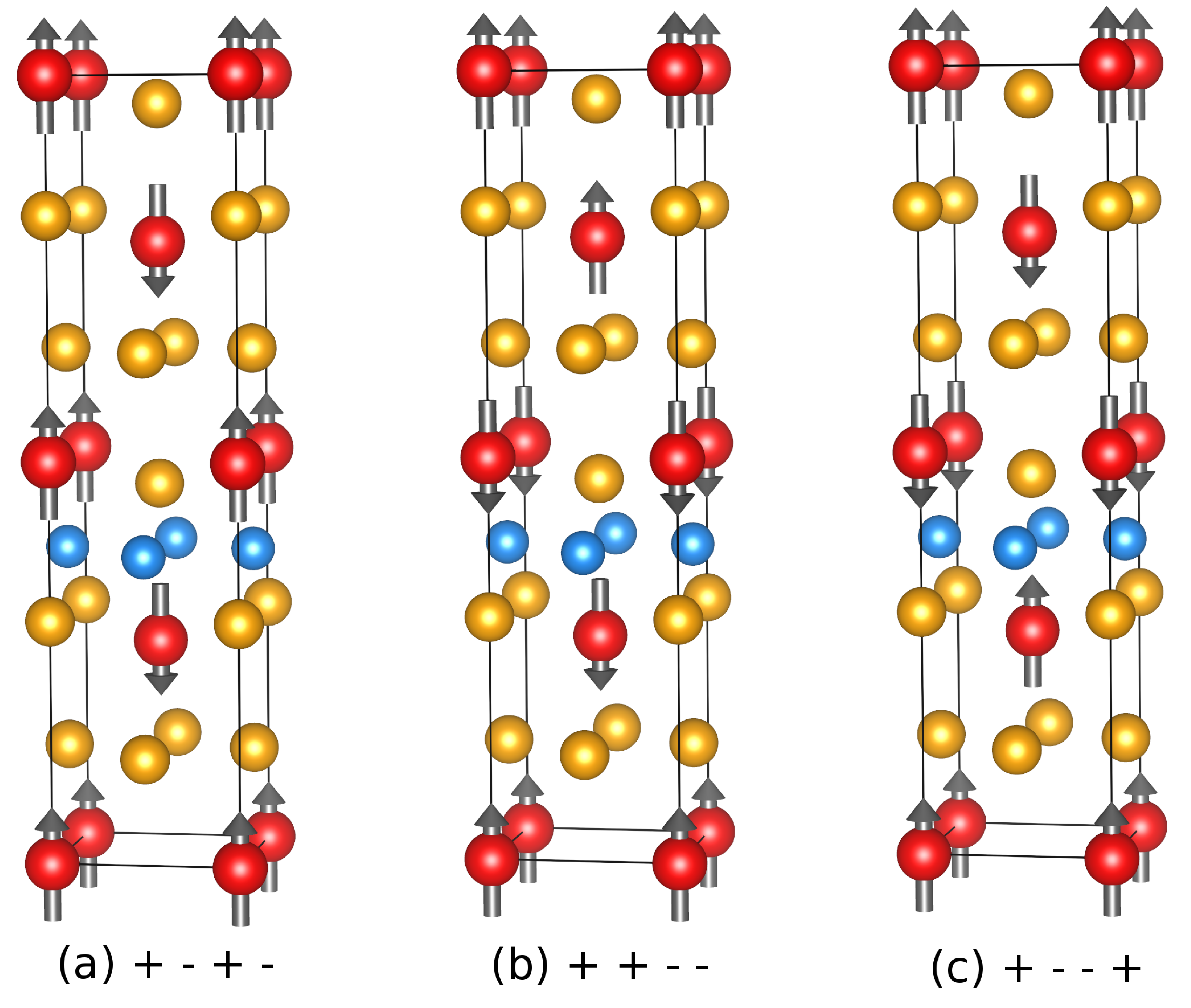}}
\caption{\label{fig:afm_configs} 
Three antiferromagnetic structures of UNi$_{0.5}$Sb$_2$ based on the proposed supercell model with occupation deficiency.
The considered uniaxial antiferromagnetic configurations are (a) $+-+-$, (b) $++--$, and (c) $+--+$.
}
\end{figure}
The presented approach of structure modeling is supported by the results of calculated magnetic moments.
For the proposed supercell model of UNi$_{0.5}$Sb$_2$ we consider three uniaxial antiferromagnetic configurations: $+-+-$, $++--$, and $+--+$, see Fig.~\ref{fig:afm_configs}.
The calculations with PBE approximation and spin-orbit coupling lead to the lowest total energy for $+-+-$ antiferromagnetic structure and the $++--$ and $+--+$ configurations are 49 and 27~meV~per~unit~cell higher in total energy, respectively.
For the unfavorable $++--$ configuration the two layers of U atoms adjacent to the layer of Ni are oriented ferromagnetically, thus induce a small magnetic moment of 0.08~$\mu_{\mathrm{B}}$ on Ni atoms.

\begin{table}
\caption{\label{tab:mag_mom} 
The calculated spin ($m_s$), orbital ($m_l$), and total ($m_j$) magnetic moments on uranium atoms (U$_1$, U$_2$, U$_3$) for USb$_2$, UNi$_{0.33}$Sb$_2$, UNi$_{0.5}$Sb$_2$, UNi$_{0.66}$Sb$_2$, and UNiSb$_2$ compounds. 
The corresponding magnetic moments averaged over the contributing uranium atoms are also presented. 
Two presented sets of results are obtained with the Perdew--Burke--Ernzerhof (PBE) exchange-correlation potential.
For the second set we included an additional orbital polarization (OP) for U~5$f$ and Ni~3$d$ orbitals.
\vspace{2mm}
}
%\rotatebox{90}{
\begin{tabular}{|c|cc|rrrrr|}      % Alignment for each cell: l=left, c=center, r=right
\hline \hline
&	&$x$	&0	&0.33	&0.5	&0.66	&1\\
\hline
&	&$m_s$	&1.56	&1.67	&1.68	&1.63	&1.63\\
&U$_1$	&$m_l$	&-2.59	&-2.68	&-2.68	&-2.20	&-2.19\\
&	&$m_j$	&-1.03	&-1.00	&-1.00	&-0.57	&-0.56\\
\cline{2-8}	
&	&$m_s$	&	&1.63	&1.63	&1.66	&\\
\multirow{3}{*}{\rotatebox[origin=c]{90}{PBE}}& U$_2$	&$m_l$	&	&-2.24	&-2.22	&-2.66	&\\
&	&$m_j$	&	&-0.61	&-0.59	&-1.00	&\\
\cline{2-8}	
&	&$m_s$	&	&1.65	&	&1.63	&\\
&U$_3$	&$m_l$	&	&-2.67	&	&-2.21	&\\
&	&$m_j$	&	&-1.02	&	&-0.58	&\\
\cline{2-8}	
&	&$\tilde m_s$	&1.56	&1.65	&1.65	&1.64	&1.63\\
& 	&$\tilde m_l$	&-2.59	&-2.53	&-2.45	&-2.36	&-2.19\\
&	&|$\tilde m_j$|&1.03	&0.88	&0.79	&0.72	&0.56\\
\hline
%&	&$x$	&0	&0.33	&0.5	&0.66	&1\\
%\cline{2-8}	
&	&$m_s$	&1.79	&1.88	&1.89	&1.88	&1.85\\
&U$_1$	&$m_l$	&-3.49	&-3.59	&-3.59	&-3.53	&-3.46\\
&	&$m_j$	&-1.70	&-1.72	&-1.70	&-1.65	&-1.61\\
\cline{2-8}	
\multirow{3}{*}{\rotatebox[origin=c]{90}{PBE+OP}}&	&$m_s$	&	&1.88	&1.87	&1.86	&\\
&U$_2$	&$m_l$	&	&-3.53	&-3.53	&-3.56	&\\
&	&$m_j$	&	&-1.65	&-1.66	&-1.70	&\\
\cline{2-8}	
&	&$m_s$	&	&1.87	&	&1.86	&\\
&U$_3$	&$m_l$	&	&-3.59	&	&-3.48	&\\
&	&$m_j$	&	&-1.72	&	&-1.62	&\\
\cline{2-8}	
&	&$\tilde m_s$	&1.79	&1.87	&1.88	&1.86	&1.85\\
& 	&$\tilde m_l$	&-3.49	&-3.57	&-3.56	&-3.52	&-3.46\\
&	&|$\tilde m_j$|&1.70	&1.69	&1.68	&1.66	&1.61\\
\hline \hline   
 \end{tabular}
% }
\end{table}

\begin{figure}
\centering
{\includegraphics[trim = 0 0 0 0,clip,width=1\columnwidth]{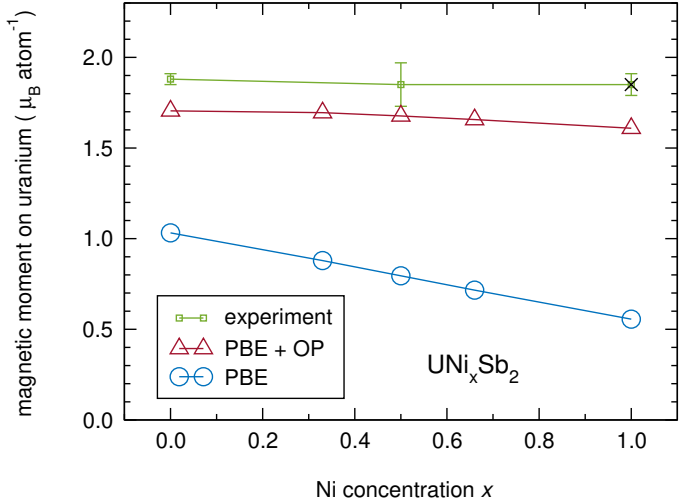}}
\caption{\label{fig:mj_vs_x} 
The calculated dependence of average total magnetic moment on uranium atom \textit{versus} Ni concentration $x$ for UNi$_{x}$Sb$_2$ system. 
The corresponding experimental results for USb$_2$~\cite{tsutsui_hyperfine_2004}, UNi$_{0.5}$Sb$_2$~\cite{torikachvili_structural_2011}, and an isostructural (marked with cross) UNiAs$_2$~\cite{fischer_antiferromagnetism_1989} are presented for comparison. 
Two sets of results are obtained with the Perdew--Burke--Ernzerhof (PBE) exchange-correlation potential, for one of them we additionally included the orbital polarization (OP).
}
\end{figure}

We calculate also the magnetic moments for the antiferromagnetic structures of all considered UNi$_{x}$Sb$_2$ structures ($x$~=~0, 1/3, 1/2, 2/3, 1).
The calculated spin ($m_s$), orbital ($m_l$), and total ($m_j$) magnetic moments on uranium atoms are presented in Tab.~\ref{tab:mag_mom}.
One set of results (PBE) is based on the simple PBE approximation,
another set (PBE+OP) is obtained with the additional potential called orbital polarization (OP) in the form introduced by Brooks, Johansson, and Eriksson~\cite{brooks_calculated_1985, eriksson_orbital_1990} and with OP parameters calculated \textit{ab initio}.
For the both sets the spin-orbit coupling is included.
As some of the considered structures have more than one pair of antiferromagnetically coupled U atoms, we evaluate also the average values of the magnetic moments on U atoms.
The total magnetic moments on U atom |$\tilde m_j$(U)| for basic composition USb$_2$ is equal 1.03~$\mu_{\mathrm{B}}$ in the PBE approach and it comprises of spin  $\tilde m_s$(U)~=~1.56~$\mu_{\mathrm{B}}$ and orbital $ \tilde m_l$(U)~=~-2.59~$\mu_{\mathrm{B}}$ magnetic moments. 
The corresponding values obtained in the PBE+OP approach are 1.70, 1.79, and -3.49~$\mu_{\mathrm{B}}$, respectively.
In comparison to the simple PBE, the orbital polarization increases significantly the orbital contribution, simultaneously increasing the total magnetic moment.
Similar results for other compositions are collected in Tab.~\ref{tab:mag_mom}.
Figure~\ref{fig:mj_vs_x} presents the calculated dependencies of |$\tilde m_j$(U)| \textit{versus} Ni concentration $x$.
For the PBE and PBE+OP approaches we observe a decrease of the magnetic moment with $x$.
Although, for the PBE+OP the drop is much lower and the values of magnetic moment on U better correspond to the experimental results for 
USb$_2$ (1.88~\textpm~0.03~$\mu_{\mathrm{B}}$)~\cite{tsutsui_hyperfine_2004}, 
UNi$_{0.5}$Sb$_2$ (1.85~\textpm~0.12~$\mu_{\mathrm{B}}$)~\cite{torikachvili_structural_2011}, 
and an isostructural UNiAs$_2$ (1.85~\textpm~0.06~$\mu_{\mathrm{B}}$)~\cite{fischer_antiferromagnetism_1989}.
For the UNi$_{0.5}$Sb$_2$ the magnetic moment on U calculated with PBE+OP (1.68~$\mu_{\mathrm{B}}$) is smaller than the experimental one (1.85~\textpm~0.12 $\mu_{\mathrm{B}}$).~\cite{torikachvili_structural_2011}

\section{Summary and conclusions}

The problem of modeling of the layered structures with occupation deficiency was addressed with the method based on the supercells.
It can be summarized in four steps: 
a) start with a fully occupied single unit cell;  
b) multiply it along the $z$ direction;
c) remove the proper atomic layers; 
d) optimize the structures.
We proved the utility of the supercell method by predicting the five crystal structures of the UNi$_{x}$Sb$_2$ phase diagram ($x$~=~0, 1/3, 1/2, 2/3, 1).
The credibility of the method was confirmed by comparison of the evaluated $c(x)$ plot with the corresponding experimental results from literature.
Based on the obtained supercells of UNi$_{x}$Sb$_2$ and including the orbital polarization corrections we also calculated the magnetic moments on uranium atoms.
The presented method of modeling can be applied to support the XRD analysis where the calculated model can be used as an initial structure to an experimental fit. 
The calculated structures of the UNi$_{x}$Sb$_2$ phase diagram can be used for further electronic structure calculations or as the basis models for the isostructural phases.

\section{Acknowledgements}

The authors would like to thank professor Peter Oppeneer for helpful discussion and comments. 
MW acknowledges the financial support from the Foundation of Polish Science grant HOMING. 
The HOMING programme is co-financed by the European Union under the European Regional Development Fund.

\end{sloppypar}

\section{Bibliography}

\bibliography{uni05sb2}        % Exported_Items.bib is the name of our database  

\end{document}